\documentclass[aps,prl,twocolumn,superscriptaddress,unsortedaddress]{revtex4-1}

\usepackage{amsmath,amssymb}
\usepackage{bm}
\usepackage[dvipdfm]{graphicx}
\usepackage{color}
\usepackage{mathrsfs}
\usepackage{amsmath}
\usepackage{natbib}
\newcommand\BaCoGeO{Ba$_2$CoGe$_2$O$_{7}$}

\begin{document}

\title{Spin nematic interaction in multiferroic compound Ba$_{2}$CoGe$_{2}$O$_{7}$}

%% Notice placement of commas and superscripts and use of &
%% in the author list

\author{M. Soda}
\affiliation{Neutron Science Laboratory, Institute for Solid State
Physics, University of Tokyo, Tokai, Ibaraki 319-1106, Japan}
\author{M. Matsumoto}
\affiliation{Department of Physics, Shizuoka University, Shizuoka 422-8529, Japan}
\author{M. M{\aa}nsson}
\affiliation{Laboratory for Quantum Magnetism (LQM), \'{E}cole Polytechnique F\'{e}d\'{e}rale de Lausanne (EPFL), Station 3, CH-1015 Lausanne, Switzerland}
\affiliation{Laboratory for Neutron Scattering, Paul Scherrer Institut, {CH-5232 Villigen PSI,} Switzerland}
\author{S. Ohira-Kawamura}
%\affiliation{Materials and Life Science Division, J-PARC Center, Tokai, Ibaraki 319-1195, Japan}
\author{K. Nakajima}
\affiliation{Materials and Life Science Division, J-PARC Center, Tokai, Ibaraki 319-1195, Japan}
\author{R. Shiina}
\affiliation{Department of Materials Science and Technology, Niigata University, Niigata 950-2181, Japan}
\author{T. Masuda}
 \email{masuda@issp.u-tokyo.ac.jp}
\affiliation{Neutron Science Laboratory, Institute for Solid State
Physics, University of Tokyo, Tokai, Ibaraki 319-1106, Japan}

\date{\today}

\begin{abstract}
We demonstrate the existence of the spin nematic interactions in
an easy-plane type antiferromagnet \BaCoGeO\ by exploring the magnetic anisotropy
and spin dynamics. Combination of neutron scattering and magnetic susceptibility
measurements reveals that the origin of the in-plane anisotropy is an antiferro-type
interaction of the spin nematic operator.
The relation between the nematic
operator and the electric polarization in the ligand symmetry of this compound
is presented. The introduction of the spin nematic interaction is useful to
understand the physics of spin and electric dipole in multiferroic compounds.
\end{abstract}


\maketitle

Symmetry breaking of time reversal and space
inversion allows spontaneous order both in magnetism and dielectricity\cite{Smolenskii}.
The enhanced simultaneous order, multiferroics\cite{Fiebig,Eerenstein},
has been extensively studied since the discovery of its experimental
realization in
the perovskite manganite TbMnO$_3$\cite{KimuraNature}.
The microscopic consideration
of electronic states taking into account a spin-orbit (SO) interaction
and symmetry of crystals reveals
 the relationship between the structures of spin ${\bm S}$
and polarization ${\bm P}$\cite{Katsura,Mostovoy,Sergienko,Jia07,Arima}.
So far, in many multiferroic materials, the stability of
complex spin orders has been focused in relation with
induced electric polarizations. On the other hand, existence
of an interaction among electric polarizations and its
interplay with the spin interaction still remains unclear.
In this context, a new material \BaCoGeO\ provides a simple
and quite interesting playground in which the polarizations
are described by
a rank two symmetric tensor of local spin operators,
reflecting a symmetry of
the tetrahedral unit involving the magnetic Co ions. One may
call the spin representation of the polarization nematic or
quadrupole, and can analyze their interaction on an equal
footing with the spin exchange interaction. The main purpose
of this paper is to give convincing evidence of realizing the
nematic interaction and to clarify its crucial role in the
magnetic anisotropy and the low
energy physics in \BaCoGeO .

\begin{figure}
\includegraphics[width=8.6cm]{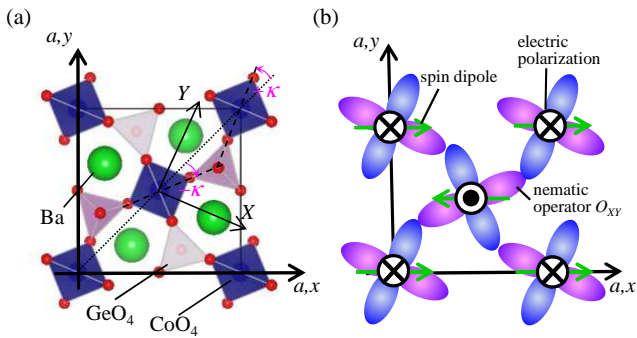}
\caption{
(a) Crystal structure of \BaCoGeO .
CoO$_4$ tetrahedra separated by GeO$_4$ tetrahedra form
a square lattice in the (001) plane.
The $x$ and $y$ axes in the global coordinate of spin
are defined along [100] and [010] directions.
The $X$ and $Y$ axes are locally defined on each CoO$_4$ tetrahedron.
(b) Structures of spin dipoles, spin nematic operator $O_{XY}$,
and electric polarizations at zero magnetic field in \BaCoGeO .
}
\label{fig1}
\end{figure}

Two-dimensional square lattice antiferromagnet \BaCoGeO\ includes
metal-ligand motif with high symmetry but without inversion center.
The crystal structure is tetragonal $P{\bar 4}2_1m$ as schematically shown in
Fig. \ref{fig1}(a).
Co$^{2+}$ ions carry spin $S=3/2$ and the CoO$_4$ tetrahedron is distorted
along the $c$ direction.
The compound exhibits an antiferromagnetic transition at
$T_{\mathrm{N}}$=6.7 K~\cite{Sato} and
a staggered antiferromagnetic
structure in the (001) plane with slight canting\cite{Zheludev}.
%The direction of the spin in the plane is rather controversial.\cite{Romhanyi,Hutanu}
Below $T_{\mathrm{N}}$, an electric polarization
induced by a magnetic field is
observed\cite{Murakawa,Murakawa12,Yi}.
The origin of the polarization was
explained by a spin-dependent $d$-$p$ hybridization
mechanism\cite{Murakawa}.

The electrons in the Co ion at the center of an isolated O$_4$ tetrahedron
suffer from a crystal field potential of the point group $D_{2d}$.
In this case, as we shall show later, conventional spin
interaction does not break the spin rotation symmetry in the plane $(001)$.
In the present study,
however, the combination of inelastic neutron
scattering with high energy resolution
and magnetic susceptibility measurements in \BaCoGeO\ reveals existence of
the distinct anisotropy
in the plane. We argue through the following study that an unconventional nematic
interaction is the origin of the anisotropy.

%method
Stoichiometric quantities of BaCO$_{3}$, CoO, and GeO$_{2}$ powders were mixed,
and sintered at 900 $^{\circ}$C for 24 h. The obtained powder is pressed into
rods and they were sintered at 1,000 $^{\circ}$C for 24 h.
Using these rods as starting materials, single crystals
were grown in air at a rate of 1 mm/h by floating zone method.
Typical dimension of the crystals is about 8 mm in diameter and
40 mm in length.
We confirmed absence of impurity phase by using powder X-ray diffractometer.
Bulk magnetization was measured using a conventional SQUID
magnetometer.
Neutron measurements were carried out using the cold-neutron
triple axis spectrometer TASP installed at SINQ/PSI, Switzerland.
Throughout this paper,
we use the tetragonal unit cell with $a$=$b$=8.410 \AA\
and $c$=5.537 \AA.
The scattering plane was the $a$ - $c$ plane.
The final neutron energy was set at 5 meV.
The magnetic fields were applied along a direction vertical
to the scattering plane (magnetic field $H$//[010]) using
a superconducting magnet. Neutron measurements were also
carried out using cold-neutron disk-chopper spectrometer
AMATERAS of direct geometry type installed at J-PARC, Japan.
The chopper condition was set so that the incident neutron energies $E_{\rm i}$
was 3.14meV and the resolutions of the energy transfer
at the elastic position was 0.075meV.

%\section{Results}

\begin{figure}
\includegraphics[width=8.6cm]{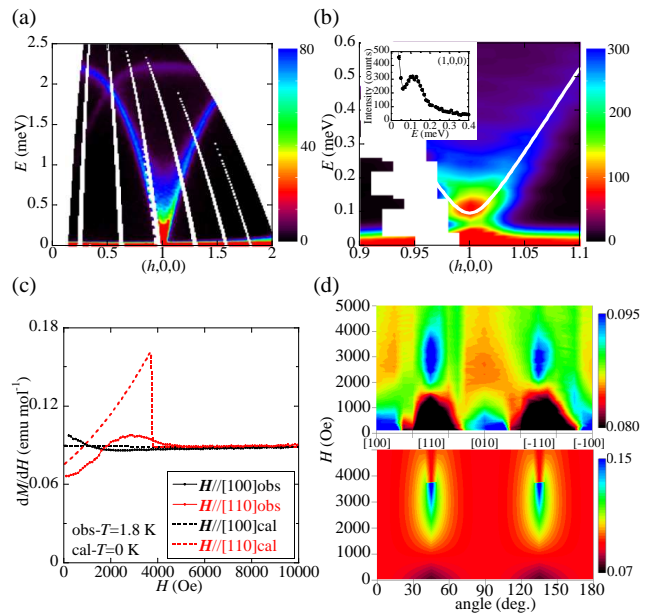}
\caption{
(a) Inelastic neutron scattering spectrum obtained by
using AMATERAS spectrometer in J-PARC.
Well-defined spin-wave excitations are observed.
(b) Inelastic neutron scattering spectrum in low energy range.
The calculated dispersion is shown by
the white solid curve.
The inset shows the constant-Q scan at
{\bf Q}$=(1,0,0).$
(c)
Bulk susceptibility ${\rm d}M/{\rm d}H$ measured at $T$ = 1.8 K.
Red and black symbols indicate the ${\rm d}M/{\rm d}H$
for $H \parallel [110]$ and $H \parallel [100]$, respectively.
Field cool processes are shown.
Red and black dashed curves indicate the ${\rm d}M/{\rm d}H$ at $T$ = 0 K
calculated by mean-field theory.
(d)
The measured angular dependence of ${\rm d}M/{\rm d}H$ in upper panel
and the calculated ones in lower panel.
}
\label{fig2}
\end{figure}

%{\bf Magnetic anisotropy in \BaCoGeO .}
In Fig. \ref{fig2}(a) the neutron spectrum is shown and
a couple of spin-wave modes with the band energy of
2.2 meV are observed.
Focused on the low energy range
a clear anisotropy gap of approximately
0.10 meV at the
antiferromagnetic zone center  ${\bm Q} = (1,0,0)$
is observed in Fig. \ref{fig2}(b).
The magnetic susceptibilities ${\rm d}M/{\rm d}H$ in field applying
along $[100]$ and $[110]$ directions are shown in Fig. \ref{fig2}(c).
In the latter a peak due to a spin flop (SF) transition is observed at $H \sim$ 3 kOe
that is consistent with the energy scale of the anisotropy gap,
while in the former the SF transition is absent.
The angular dependence of ${\rm d}M/{\rm d}H$ is summarized
in the upper panel of Fig. \ref{fig2}(d), where
four-fold rotational symmetry with the enhanced SF field at $[110]$
and $[-110]$ is observed.
The presence of the four-fold in-plane anisotropy is consistent with the prediction
of the anisotropy by investigating the crystallographic symmetry\cite{Perez,Bordacs}.

In general the origin of the magnetic anisotropy gap in the excitation spectrum
is a single-ion anisotropy or a two-ions anisotropy.
The single-ion anisotropy in the spin state in this material
is represented
by the form $D(S^z)^2$
where the $z$-axis is the crystallographic $c$-axis.
We note that there is no $E$ term in the tetragonal system.
The sign of $D$ is believed to be positive since the direction
of spins in the ordered state is perpendicular to the $c$-axis.
Spin operators with 4th or higher order is inactive for S=3/2 in the local symmtery
and, thus, the
single ion anisotropy cannot lead to any in-plane anisotropy.
Consider next the non-local symmetry of Co ions located on the 2$a$ Wyckoff positions in the
space group $P{\bar 4}2_1m$ and the allowed two-ion
anisotropy represented by a spin-dipole interaction is an $XXZ$-type one
$J^{x}(S_{1}^{x}S_{2}^{x}+S_{1}^{y}S_{2}^{y})+J^{z}S_{1}^{z}S_{2}^{z}$.
The anisotropy is, again, does not break the rotational symmetry in the $c$-plane
nor induce the anisotropy gap.
Among the two-ion anisotropies that break the rotational symmetry,
the one having the lowest order
is represented by the interaction between spin-nematic (quadrupole) operators
, of which the exact formula will be obtained later.
This means that in \BaCoGeO\ the spin-nematic interaction is the
leading term for rotational symmetry broken
and, hence, the origin of the observed anisotropy gap is
the spin-nematic interaction.

The representation of spin-nematic interaction is obtained by considering
the electric polarization interaction.
In \BaCoGeO\ the spin nematic operator becomes equivalent to the polarization
owing to the lack of the inversion symmetry in the $D_{2d}$ point
group symmetry of the CoO$_4$ tetrahedron\cite{Romhanyi11,Penc}.
Thus, $P^X = -K_{ab}O_{YZ},$
$P^Y = -K_{ab}O_{ZX},$ and
$P^Z = -K_{c}O_{XY}$,
where $K_{ab}$ and $K_c$ are constants and
$X$, $Y$, and $Z$ are the local coordinates on CoO$_4$ tetrahedron
as shown in Fig. \ref{fig1}(a)\cite{note}.
Here the spin nematic operator is defined as $O_{\alpha \beta}=S^{\alpha}S^{\beta}+S^{\alpha}S^{\beta}$, where
$\{\alpha, \beta \} = \{X, Y, Z \}$.
%These representations can be also obtained by using the relation
%between the polarization ${\bm P}$, spin moment ${\bm S}$, and the position vector of ligands ${\bm e}_i$,
%$P \prop \sum _i ({\bm S}\cdot {\bm e}_i)^2{\bm e}_i$.
Let us consider the electrostatic polarization interaction
on the basis of these operators.
Among these, $O_{YZ}$ and $O_{ZX}$ become zero under
the spin structure aligned in the $ab$-plane, while $O_{XY}$
is finite. Then, only the $Z$ component, $P^{Z} \propto O_{XY}$, is active
and an intersite interaction of $P^Z$ is converted to a nematic
interaction of $O_{XY}$ with
an effective coupling strength
$J_p^{\rm eff}$\cite{Romhanyi11}. The explicit form is
\begin{equation}
\mathcal{H}_p = -J_p\sum_{i,j}P^Z(i)P^Z(j) = -J_p^{\rm eff}\sum _{i,j}O_{XY}(i)O_{XY}(j).
\end{equation}
Thus, intersite polarization interaction gives rise to the interaction
between the spin nematic operators.
After the transformation to the global $xy$ coordinates,
we obtain
\begin{equation}
O_{XY} = \left( \cos (2\kappa )O_{xy} - \sin (2\kappa )O_2^2 \right),
\end{equation}
where $O_{xy} = S^xS^y+S^yS^x$ and
$O_2^2 = (S^x)^2-(S^y)^2$\cite{Romhanyi11,Miyahara}.
The definition of the angle ${\kappa}$ is described in Fig. \ref{fig1} (a).
Calculation of the classical energy including antiferromagnetic spin interaction
and the nematic interaction between $O_{XY}$ operators result in a ground state
with staggered spin structure with four-fold rotational symmetry.
Thus the existence of the nematic interaction
explains the biaxial magnetic anisotropy in our experiment.

The sign of $J_p$ and the magnetic-easy axes are
determined from the angular dependence of the SF
(spin flop) field $H_{\rm SF}$.
In zero field there are four magnetic domain states; two of them share the magnetic easy axis
and so do the rest two. The direction of the axis is orthogonal to each other.
If the field is applied along one of the easy axes the spins in a
domain of which the easy axis is parallel to the field flip instantly
and those in another domain stay as they are.
Hence the value of $H_{\rm SF}$ is zero or strongly suppressed.
As the field direction tilts from the easy axes the Zeeman energy required for
the SF increases and consequently $H_{\rm SF}$ is enhanced.
Experimentally the minimum of $H_{\rm SF}$ is observed at $H \parallel [010]$
and $[100]$ and this means
that the easy axes are $[010]$ and $[100]$.
This magnetic anisotropy leads the negative sign of $J_p$ in Eq. (1),
{\it i.e.}, the antiferro-type nematic interaction that is
equivalent to the antiferroelectric polarization.
It is consistent with the zero polarization along $Z$ direction at $H$ = 0 Oe
previously reported\cite{Murakawa,Murakawa12}.
Indeed the calculation of the polarization curve assuming antiferroelectric
polarization was consistent with the experiment\cite{Romhanyi11}.
The structure of spin dipole, spin nematic operator, and
the electric polarization is depicted in Fig. \ref{fig1}(b).

To estimate the spin nematic interaction $J_p^{\rm eff}$,
we performed the calculation of magnetic dispersion relation
and magnetic susceptibility as displayed in Figs. \ref{fig2}(b), (c) and (d).
Main magnetic parameters including exchange constant and single ion anisotropy
along $z$-direction are separately obtained that will be explained later.
The dispersion is reproduced by the main parameters plus
$J_p^{\rm eff} = -0.198 {\mu}{\rm eV}$
as shown by white solid curve.
%Dzyaloshinskii-Moriya interaction
%$D_{\rm DM}^z(i,j)(S_i^xS_j^y-S_i^yS_j^x)$ is
%also included\cite{Romhanyi11} to lift the degeneracy of the magnetic structure
%and to reproduce the weak-ferromagnetic component at $H$ = 0 T,
%though it hardly affect the dispersion.
The calculated susceptibility by using the same parameters
in the neutron spectrum is shown in the lower panel of Fig. \ref{fig2}(d)
and it reasonably reproduces the experiment.
Hence the magnetic anisotropy observed both in the neutron spectrum
and bulk magnetic susceptibility is
consistently explained by introducing the antiferro-nematic spin interaction.

\begin{figure}
\includegraphics[width=8.6cm]{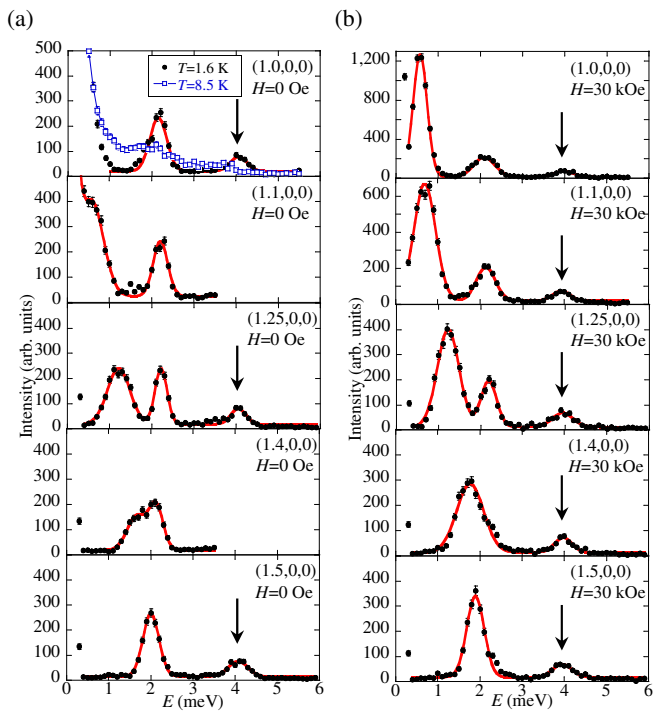}
\caption{
Typical constant-Q scans at
$H$ = 0 Oe and 30 kOe obtained by using TASP spectrometer in PSI.
Filled circles are the data at $T$ = 1.6 K and
open squares are those at $T$ = 8.5 K.
Red solid curves are fits to the data by Gaussian functions.
}
\label{fig3}
\end{figure}

\begin{figure}
\includegraphics[width=8.6cm]{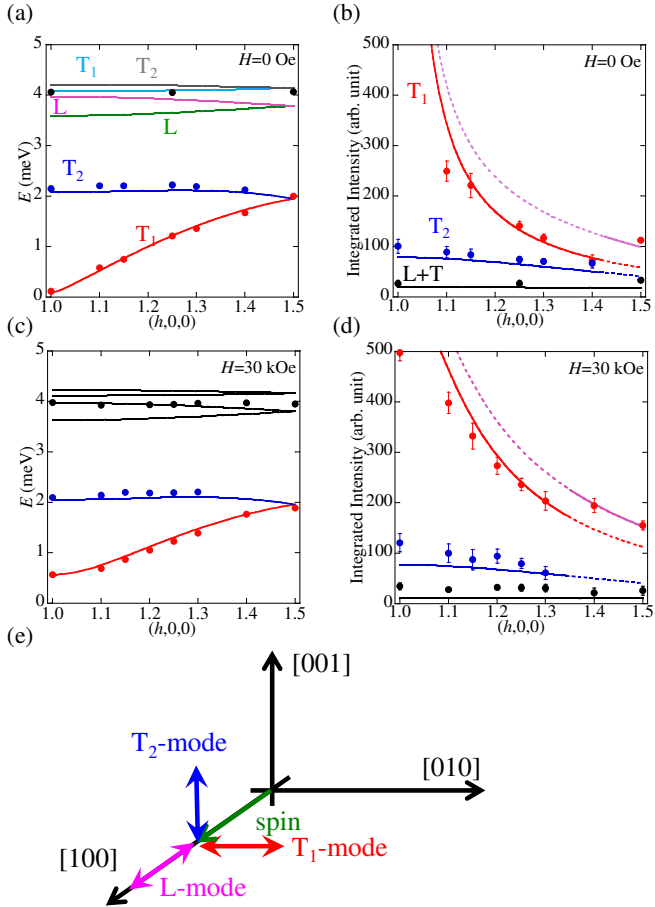}
\caption{
(a) Magnon dispersion relation for ${\bf Q}$=($h$,0,0) at $H$ = 0 Oe.
(b) Intensity of the inelastic neutron scattering at $H$ = 0 Oe.
(c),(d) Magnon dispersion and intensity for $H$=30 kOe. In each figure
the theoretical calculations are indicated by curves and
the experimental data are indicated by filled circles.
The pink curves in (b) and (d) represent the added
intensities of the lowest (red) and the second-lowest (blue) modes
for the data of which the energies are close to each other.
Similarly the black curves in (b) and (d)
are the results after adding all contributions from the
four high-energy modes around 4 meV.
(e)
Schematics of spin fluctuation for each mode at 0 Oe.
Transverse fluctuation in the $a$ - $b$ plane is named T$_1$ mode,
those along the $c$ direction is named T$_2$ mode, and
longitudinal fluctuation is named L mode.
}
\label{fig4}
\end{figure}

To determine the main magnetic parameters,
inelastic neutron scattering spectra for extended energy transfers were measured
as shown in Fig. \ref{fig3}.
In addition to the low energy excitations at
0$\lesssim \hbar \omega \lesssim$2 meV reported in
a previous neutron scattering study\cite{Zheludev},
a new excitation at $\hbar \omega$ $\sim$ 4 meV
is observed, which
corresponds to an electric-field active mode confirmed by
electromagnetic wave absorption experiments\cite{Kezsmarki,Miyahara}.
The excitation energies obtained by the Gaussian fits
to the constant-Q scans
are plotted in Figs. \ref{fig4}(a) and (c).
Obtained intensities for the excitations
are also plotted in Figs. \ref{fig4}(b) and (d).
While the lower energy modes at $\hbar \omega \lesssim$ 2 meV
are explained by a classical spin-wave theory for
square-lattice antiferromagnet,
the higher energy modes cannot be reproduced.
To analyze all the modes we use extended spin-wave theory\cite{Shiina}
that is equivalent to the bond operator formulation\cite{Sachdev,Penc,Romhanyi}.
We consider the Hamiltonian
\begin{equation}
\mathcal{H}=\sum_{i}\mathcal{H}_{\rm intra}(i)+\sum_{<i,j>}\mathcal{H}_{\rm inter}(i,j)
\end{equation}
where $\mathcal{H}_{\rm intra}(i)$ and $\mathcal{H}_{\rm inter}(i,j)$
are Hamiltonians for intrasite and intersite parts defined by
\begin{eqnarray}
\mathcal{H}_{\rm intra}(i) &=& D(S_{i}^{z})^{2} -
\sum_{\alpha=x,y,z}g^{\alpha}\mu_{B}H^{\alpha}S_{i}^{\alpha} \\
\mathcal{H}_{\rm inter}(i,j)&=&\sum_{\alpha=x,y,z}
J^{\alpha}S_{i}^{\alpha}S_{j}^{\alpha}+D_{\rm DM}^z(i,j)(S_i^xS_j^y-S_i^yS_j^x) \nonumber \\
&+& \mathcal{H}_p
\end{eqnarray}
Here, $S_{i}^{\alpha}$ is an $\alpha(=x, y, z)$ component of
the $S$ = 3/2 spin operator at the $i$-th site.
In the presence of large $D$,
the local energy eigenstates are split into two doublets of $S^z=\pm 1/2$ and $\pm 3/2$.
For $D>$0, $S^{z}$ = $\pm$1/2 doublet is stabilized and through intersite interaction
an effective $S=1/2$ spin wave is induced in the low energy range.
Further, crystal field like excitation from $S^z=\pm 1/2$ to $\pm 3/2$
exists at higher energy.

As the result of neutron
cross section analysis, we obtain the parameters summarized in
Table \ref{parameters}, which are semi-quantitatively consistent with
those obtained from ESR studies\cite{Penc}.
The spin system is approximately identified as $S=3/2$
square-lattice antiferromagnet
with large positive $D$.
Calculated
dispersion of the magnetic excitation along
$\mbox{\boldmath $Q$}$ = ($h$,0,0) for
$H$=0 Oe and 30 kOe are shown in Figs. \ref{fig4}(a) and (c)
by solid curves.
The lowest- and second lowest-lying modes are basically
connected to the interacting lowest-lying doublets of the $S^z = \pm 1/2$.
The former is the transverse fluctuation in the $a$-$b$ plane,
T$_{1}$-mode and the latter are those in
the $c$-direction, T$_{2}$-mode as depicted in Fig. \ref{fig4}(e).
On the other hand, the mode around 4 meV comes
from the higher-lying doublets and contains
a large magnitude of longitudinal fluctuations of the ordered moment,
L-mode \cite{Penc,Romhanyi} and
a small magnitude of transverse fluctuations, T$_1$- and T$_2$-modes.
These L, T$_1$, and T$_2$ modes are consistent with those
identified in the previous ESR studies\cite{Penc}.

%\section{Discussion}

The existence of spin nematic interaction in multiferroic
compound \BaCoGeO\ was revealed for the first time  by a combination of
inelastic neutron scattering and bulk magnetization measurements.
Several key features in this compound make the physics very interesting;
strong SO interaction, the metal-ligand symmetry with no inversion center,
and absence of conventional
in-plane anisotropy.
In the course of this study, we have emphasized that such circumstances result in
a characteristic correlation among spin nematic, spin dipole, electric polarization,
which is totally described in the spin basis.
Using this framework, we succeeded in determining the dielectric energy
by measuring precisely the magnetic anisotropy energy.

From magnetic point of view the electric
field tunes the direction of magnetic moment and
the determined dielectric energy
can be a performance index for multiferroicity. Further search for magnets that exhibit
small magnetic anisotropy, and therefore small electric field for achieving
ferroelectricity, would lead a practical multiferroic device that
switches the direction of spin moment by a small electric field.

%In addition, a cover letter needs to be written with the
%following:
%\begin{enumerate}
% \item A 100 word or less summary indicating on scientific grounds
%why the paper should be considered for a wide-ranging journal like
%\textsl{Nature} instead of a more narrowly focussed journal.
% \item A 100 word or less summary aimed at a non-scientific audience,
%written at the level of a national newspaper.  It may be used for
%\textsl{Nature}'s press release or other general publicity.
% \item The cover letter should state clearly what is included as the
%submission, including number of figures, supporting manuscripts
%and any Supplementary Information (specifying number of items and
%format).
% \item The cover letter should also state the number of
%words of text in the paper; the number of figures and parts of
%figures (for example, 4 figures, comprising 16 separate panels in
%total); a rough estimate of the desired final size of figures in
%terms of number of pages; and a full current postal address,
%telephone and fax numbers, and current e-mail address.
%\end{enumerate}

%% Put the bibliography here, most people will use BiBTeX in
%% which case the environment below should be replaced with
%% the \bibliography{} command.

% \begin{thebibliography}{1}
% \bibitem{dummy} Articles are restricted to 50 references, Letters
% to 30.
% \bibitem{dummyb} No compound references -- only one source per
% reference.
% \end{thebibliography}

%\bibliographystyle{apsrev}

%% Here is the endmatter stuff: Supplementary Info, etc.
%% Use \item's to separate, default label is "Acknowledgements"

\begin{acknowledgments}
Prof. A. Zheludev and Dr. S. Gvasaliya are greatly appreciated for facilitating the
neutron scattering experiment and for fruitful discussion.
This work was supported by KAKENHI (24340077, 20740171, 24740224, and
23540390).
\end{acknowledgments}

%%
%% TABLES
%%
%% If there are any tables, put them here.
%%

\begin{table}
\centering
\caption{Parameters obtained by extended spin wave calculations.}
\label{parameters}
\medskip
 \begin{tabular}{l l l l l l l}
$J^{x}$, $J^{y}$~(meV) &  $J^{z}$~(meV) & $D$~(meV)
& $J_p^{\rm eff}$~($\mu$eV)&$D_{\rm DM}^z$~($\mu$eV)  \\
 \hline
0.208 & 0.253 & 1.034 & -0.198 & 8.61  \\
 \hline
 \end{tabular}
\end{table}

%\bibliography{sample}

\begin{thebibliography}{23}
\bibitem{note}
These representations can be also obtained by using the relation
between the polarization {\bf P}, spin moment {\bf S}, and the
position vector of ligands {\bf e}$_i$,
{\bf P}$ \propto \Sigma _i (${\bf S}$\cdot ${\bf e}$_i)^2${\bf e}$_i$.
\end{thebibliography}

\end{document}